\documentclass{article}
\usepackage{fleqn,amssymb,amsmath}
\newtheorem{thm}{Statement}
\def\stackunder#1#2{\mathrel{\mathop{#2}\limits_{#1}}}
\def\stackunder#1#2{\mathrel{\mathop{#2}\limits_{#1}}}

\newcommand{\Const}{\mathop{\rm Const}\nolimits}

\begin{document}

\begin{center}
{\Large Instability Model of the Universe with De Sitter Beginning}\\
Yurii Ignat'ev\\
Lobachevsky Institute of Mathematics and Mechanics, Kazan Federal University,\\ Kremleovskaya str., 35, Kazan 420008, Russia
\end{center}

The physical instability of the Universe model with de Sitter beginning is proved in this article. 1. It is shown that even a small addition of ultrarelativistic matter turns the de Sitter Universe into the Universe with finite past. 2. Constant solution of equation of model with constant scalar field is shown to be instable. 3. It is also shown that lateral gravitational perturbations of such model make it instable near of cosmological singularity.
The conclusion is made that de Sitter stage of the Universe evolution most likely should be preceded by the ultrarelativistic stage.\\
Keywords: De Sitter universe, instability, finite lifetime

\section*{Introduction}
A great number of articles is devoted to models of Universe with de Sitter beginning, i.e. models with {\it primordial} inflation. By primordial inflation we will mean such a situation where inflation exists from the instant of cosmological singularity. The analysis of {\it mathematical model} of the early Universe reveals a series of its defects among which there are quite serious ones prejudicing its mathematical correctness. The article is devoted to this analysis.

\section{The Analysis of Minimal Model}
\subsection{Equations of Minimal Model}
As is well known\footnote{see e.g. \cite{Land_Field}}, Einstein equations in case of homogenous isotropic cosmological model with zero 3-dimensional curvature
\begin{eqnarray}\label{Freedman}
ds^2=dt^2-a^2(t)(dx^2+dy^2+dz^2)\nonumber\\ \equiv a^2(\eta)(d\eta^2-dx^2-dy^2-dz^2);\\
\label{t-eta}
t=\int a(\eta)d\eta \Leftrightarrow \eta=\int\frac{dt}{a(t)}
\end{eqnarray}
are reduced to system of two ordinary first-order differential equations
\footnote{Planck system of units is used everywhere: $\hbar=G=c=1$}:
\begin{equation}\label{einst}
\frac{\dot{a}^2}{a^2}=\frac{8\pi}{3}\varepsilon;
\end{equation}
\begin{equation}\label{energy_cons}
\dot{\varepsilon}+3\frac{\dot{a}}{a}(\varepsilon+{\rm p}(\varepsilon))=0.
\end{equation}

Let:
\begin{equation}\label{summmary}
\varepsilon=\varepsilon_{m}+\varepsilon_{s};\quad {\rm p}={\rm p}_{m}+{\rm p}_{s},
\end{equation}
where $\varepsilon_{m},{\rm p}_{m}$ is an energy density and matter pres\-sure, $\varepsilon_{s},{\rm p}_{s}$ is an energy density and pressure of all sorts of fundamental fields including scalar ones leading to the Universe acceleration.  Let us accept minimal model of interaction where
 these two com\-po\-nents are conserved independently:
\begin{eqnarray}\label{energy_{s}}
\dot{\varepsilon}_{s}+3\frac{\dot{a}}{a}(\varepsilon_{s}+{\rm p}_{s}(\varepsilon_{s}))=0,\\%
\label{energy_{m}}
\dot{\varepsilon}_{m}+3\frac{\dot{a}}{a}(\varepsilon_{m}+{\rm p}_{m}(\varepsilon_{m}))=0.
\end{eqnarray}
Let then:
\begin{eqnarray}\label{kappa_{s}}
p_{s}=-\varepsilon_{s};& \\
\label{kappa_k}
p_{m}=\kappa_{m} \varepsilon_{m};& -1<\kappa_{m} \leq 1.
\end{eqnarray}
Integrating equations (\ref{energy_{s}}), (\ref{energy_{m}}), we obtain:
\begin{eqnarray}\label{es}
\varepsilon_{s}=\varepsilon^0_{s}={\Const};\\
\label{em}
\varepsilon_{m}=Ca^{-3(\kappa_{m}+1)}.
\end{eqnarray}

\subsection{Exact Cosmological Solution for a Case With Addition Ultrarelativistic Component}
Let us consider a case of ultrarelativistic matter
\begin{equation}\label{em1/3}
p_{m}=\frac{1}{3}\varepsilon_{m}
\end{equation}
and positive density of dark matter
\begin{equation}\label{es>0}
\varepsilon^0_{s}>0.
\end{equation}
Substituting (\ref{es}) and (\ref{em}) with an account of (\ref{summmary}) into Einstein equations (\ref{einst}):
\begin{equation}\label{einst1}
\frac{\dot{a}}{a}=\sqrt{\frac{8\pi}{3}(\varepsilon_{s}+\varepsilon_{m})}
\end{equation}
and integrating it for a case of ultrarelativistic matter, we
find the solution of Einstein equations \cite{Yu_LTE}:
\begin{eqnarray}\label{solv1/3}
a^2(\tau)=&a^2(\tau_0)\cosh(\tau-\tau_0)+\nonumber\\
 & \sqrt{a^4(\tau_0)+\gamma^2}\sinh(\tau-\tau_0),
\end{eqnarray}
where $\tau$ is a dimensionless physical time
\begin{equation}\label{tau}
\tau=\frac{t}{t_c}\equiv t\sqrt{\frac{32\pi}{3}\varepsilon^0_{s}},
\end{equation}
and constant $\gamma^2$ is equal to ratio of radiant energy density to dark matter energy density at the instant when scale factor is equal to one:
\begin{equation}\label{gamma}
\gamma^2=\frac{C}{\varepsilon^0_{s}}\equiv \left.\frac{\varepsilon_{m}(a)}{\varepsilon^0_{s}}\right|_{a=1}.
\end{equation}

Let us put for definiteness (with an account of arbitrariness of choice of a scale unit, i.e., $a_0$):
\begin{equation}\label{const}
\tau_0=0;\quad a(\tau_0)=1
\end{equation}
and simplify expression (\ref{solv1/3}) to form:
\begin{equation}\label{solv1/3_0}
a^2(\tau)=\cosh(\tau)+\sqrt{1+\gamma^2}\sinh(\tau).
\end{equation}
At $\tau\to+\infty$ and $\gamma\to0$ solution (\ref{solv1/3_0}) becomes an inflation one:
\begin{equation}\label{infl}
a(\tau)\approx \mathrm{e}^\tau.
\end{equation}

Let us investigate the behavior of scale factor (\ref{solv1/3_0}) at great negative values of time
\begin{equation}\label{limit_T}
\tau\to -\infty.
\end{equation}
Finding the asymptotics, we obtain:
\begin{equation}\label{lim_a}
a^2(\tau)\stackunder{\tau\to-\infty}{\simeq} \frac{1}{2}{\rm e}^\tau\bigl(1-\sqrt{1+\gamma^2}\bigr),
\end{equation}
where from it is obvious that only at strict equality:
\begin{equation}\label{g0}
\gamma\equiv 0\Rightarrow \varepsilon_{m}\equiv 0
\end{equation}
we obtain in extreme case
\begin{equation}\label{a-8}
a(-\infty)= 0.
\end{equation}
At non-zero values of $\gamma$ resolution of relation (\ref{solv1/3_0}) with respect to $\tau$, let us find the non-negativity constraint $a^2(\tau)$:
\begin{eqnarray}\label{T8}
\tau&\geq &\tau_{-\infty}, \nonumber \\
\tau_{-\infty}&=& -\frac{1}{2}\ln\frac{\sqrt{1+\gamma^2}+1}{\sqrt{1+\gamma^2}-1}<0,
\end{eqnarray}
so that:
\begin{equation}\label{a=0}
a(\tau_{-\infty})=0.
\end{equation}
In particular, at small values of constant $\gamma$ we find from here:
\begin{equation}\label{tg8}
\tau_{-\infty}\backsimeq -\ln\frac{1}{\gamma}.
\end{equation}
Exactly this instant $a(\tau_{-\infty})=0$ corresponds to cosmological singularity.
Therefore even at quite small ultrarelativistic ``additives'' to
``pure vacuum state'' , $\gamma\sim 10^{-6}$ we discover ``infinitely distant past''
to be not that distant: $\tau_{-\infty} \thicksim-6$.
Proceeding to timing $t$ from the instant of cosmological sin\-gu\-la\-rity:
\begin{equation}\label{t-t8}
\tau=\tau_{-\infty}+\frac{t}{t_c},
\end{equation}
we transform solution (\ref{solv1/3_0}) to simple form:
\begin{equation}\label{exact_8}
a^2(t)=\gamma \sinh(t/t_c).
\end{equation}
At small cosmological times $t/t_c\ll 1$ we obtain from here an ultrarelativistic asymptotics near singu\-la\-rity point:
\begin{equation}\label{deltat}
 a(\tau_{-\infty}+t/t_c) =\sqrt{\gamma \frac{t}{t_c}}\equiv\left(\frac{32\pi\varepsilon_m}{3}\right)^{1/4}t^{1/2}.
\end{equation}
From (\ref{exact_8}) it follows that the exact solution comes to inflationary mode at times
\begin{equation}\label{inflat_t}
t\gg t_c \Rightarrow a(t)\approx \sqrt{\gamma}\mathrm{e}^{t/t_c}.
\end{equation}

\subsection{Analysis of a Minimal Model With Arbitrary Equation of Matter State}
Let us consider Einstein equation with addition component of arbitrary nature using solutions (\ref{em}):
\begin{equation}\label{einst2}
\frac{\dot{a}^2}{a^2}=\frac{8\pi}{3}\varepsilon^0_{s}(1+\gamma a^{-3(1+\kappa_{m})}).
\end{equation}
Let us consider behavior of solutions of this equation near singularity $a\to 0$. In this case it is obvious that the dominant term in the right part of
 (\ref{einst2}) is again a matter term. Thus, near singularity we get instead of (\ref{einst2}):
\begin{equation}\label{einst3}
\frac{d\ln a}{d\tau}=\frac{1}{2}\sqrt{\varepsilon^0_{s}\gamma} a^{-\frac{3}{2}(1+\kappa_{m})}.
\end{equation}
Solving this equations, we obtain:
\begin{equation}\label{a_{m}_0}
a(\tau)=\left[1+\frac{3}{4}(\kappa_{m}+1)\gamma\tau\right]^{\frac{2}{3(\kappa_{m}+1)}}.
\end{equation}
It is thus seen that singular state $a\to0$ at chosen scale is attained at instant:
\begin{equation}\label{T-81}
\tau_\infty=-\frac{4}{3(\kappa_{m}+1)}\frac{1}{\gamma},
\end{equation}
i.e., again it is in the observed past (for mentioned parameter $\tau_{-\infty}\backsim -10^6$).

\begin{thm}
Even small additives of matter with equation of state different from inflation one ($\varepsilon+P=0$) drastically changes cosmological situation: the Universe has its beginning in the finite past and comes to inflation at later times $t\sim t_c$. In case of ultrarelativistic additive the beginning of the Universe logarithmically depends on relative portion of this matter in energy density.
\end{thm}

\section{Instability of the Standard Model of\\ Primordial Inflation}
\subsection{Standard model}
Let us consider a model of the Universe in the presence of only scalar field\footnote{Further index $(s)$ is omitted for brevity sake.}:
\begin{equation}\label{Tik}
\hskip -24pt T^{(s)}_{ik} =\frac{1}{8\pi } \left[2\Phi_{,i} \Phi _{,k} -g_{ik} \Phi _{,j} \Phi ^{,j} + m^{2} g_{ik} \Phi ^{2} \right],
\end{equation}
Let us assume for undisturbed model $\Phi(t)$. Thus we get for metrics (\ref{Freedman}):
\begin{eqnarray}
\label{Ts}
T^{(s)}_{ik}=(\varepsilon+p)\delta^4_i\delta^4_k -p g_{ik},\\
\label{e}
\varepsilon=\frac{1}{8\pi}(\dot{\Phi}^2+m^2\Phi^2);\\
\label{p}
p=\frac{1}{8\pi}(\dot{\Phi}^2-m^2\Phi^2);
\end{eqnarray}
Thus the conservation law for field  takes form:
\begin{eqnarray}\label{eq_F}
\dot{\varepsilon}+3\frac{\dot{a}}{a}(\varepsilon+{\rm p}(\varepsilon))=0 \Rightarrow\nonumber\\%
\frac{1}{4\pi}\dot{\Phi}\bigl(\ddot{\Phi}+3\frac{\dot{a}}{a}\dot{\Phi}+m^2\Phi \bigr)=0.
\end{eqnarray}
Putting in (\ref{eq_F}) $\dot\Phi\not\equiv0$, we obtain field equation:
\begin{equation}\label{Fn0}
\ddot{\Phi}+3\frac{\dot{a}}{a}\dot{\Phi}+m^2\Phi =0,
\end{equation}
This equation needs to be solved jointly with Ein\-stein equation (\ref{einst}). The system of these 2 equations comp\-rises the standard cosmological model with the only amen\-dment that instead of massive term $m^2\Phi$ in (\ref{Fn0}) it is often used an arbitrary force fun\-ction $-\nabla V$ where $V(\Phi)$ is a potential (equation (\ref{Fn0}) cor\-responds to model with a quadratic potential).

How is this system solved? Normally the fol\-lowing approximation is used (see e.g. \cite{star})
\begin{equation}\label{approxM}
m^2\ll H^2\equiv \frac{\dot{a}^2}{a^2}.
\end{equation}
In this case there must be fulfilled also the next condition:
\begin{equation}\label{approxM2}
m^2\Phi\ll \ddot{\Phi},
\end{equation}
which in normal conditions corresponds to times less than Compton ones for scalar bosons.

Thus in this approach field equation (\ref{Fn0}) is re\-duced to massless equation:
\begin{equation}\label{Fn0_m0}
\ddot{\Phi}+3\frac{\dot{a}}{a}\dot{\Phi}=0,
\end{equation}
which has its {\it particular} solution:
\begin{equation}\label{F0}
\Phi=\Phi_0=\mathrm{Const}.
\end{equation}

Herewith according to (\ref{e}) and (\ref{p}) we get:
\begin{equation}\label{p=-e}
\varepsilon=-p=\frac{1}{8\pi}m^2\Phi^2_0=\mathrm{Const}.
\end{equation}
Substituting the solution into Einstein equation (\ref{einst}) and solving it, we get an inflationary solution:
\begin{equation}\label{inflat}
a(t)=a_1 \mathrm{e}^{H_0t},\qquad H_0=\frac{m}{\sqrt{3}}|\Phi_0|.
\end{equation}
In this model the Universe beginning corresponding to cosmological singularity $a=0$ lies in the infinitely distant past:
\begin{equation}\label{begin}
a(-\infty)=0.
\end{equation}
Let us notice that Einstein equation (\ref{einst}) in the con\-si\-dered model has the following form:
\begin{equation}\label{HH0}
H^2=H^2_0,
\end{equation}
therefore due to  (\ref{inflat}) condition (\ref{approxM}) is equivalent to:
\begin{equation}\label{F>>}
|\Phi_0|\gg \sqrt{3},
\end{equation}
i.e., {\it it is fulfilled only for extremely great values of scalar field potential}.

Let us move now to the next iteration. Let us substitute Einstein equation's solution (\ref{inflat}) into the source equation of scalar field (\ref{Fn0})
\begin{equation}\label{Fn1}
\ddot{\Phi}+3H_0\dot{\Phi}+m^2\Phi=0
\end{equation}
and find its exact solution where we take into account condition (\ref{approxM}):
\begin{equation}\label{Phi1}
\Phi\approx  C_1\mathrm{e}^{- 3H_0t}+C_2 {\displaystyle\mathrm{e}^{-\frac{m^2}{3H_0}t}}.
\end{equation}
The first term in this solution attenuates rapidly with increase of cosmological time $t$\footnote{Let us notice, however, that the same term near cosmological singularity tends to infinity exponentially fast that points to instability of solution (\ref{F0}).} whereas second term remains approximately constant if
\begin{equation}\label{approx_t}
m t\ll \frac{3H_0}{m}\equiv \sqrt{3}|\Phi_0|
\end{equation}
Therefore at (\ref{approx_t}) we obtain:
\begin{equation}\label{approx_F}
\Phi\approx \Phi_0 {\displaystyle\mathrm{e}^{-\frac{m^2}{3H_0}t}}\approx \Phi_0\gg1.
\end{equation}
Let us notice that in consequence of (\ref{approxM}), condition (\ref{approx_t}) is much weaker than condition (\ref{approxM2}),
therefore inflationary solution is valid for times much greater than Compton times for scalar bosons:
\begin{equation}\label{tc}
t\ll t_c=\frac{\sqrt{3}|\Phi_0|}{m}\gg \frac{1}{m}.
\end{equation}

\subsection{Another Asymptotic Solution of Equation (\ref{Fn0})}
Let us note that we can find an another asymp\-totic solution of equation (\ref{Fn0})to an approximation (\ref{approxM2}):
\begin{eqnarray}
\ddot{\Phi}+3\frac{\dot{a}}{a}\dot{\Phi} =0\Rightarrow \nonumber\\
\label{S_Eq_m0}
a^3\dot{\Phi}=\mathrm{Const}\Rightarrow \dot{\Phi}=C/a^3(t),
\end{eqnarray}
putting now
\begin{equation}\label{C}
C\not=0.
\end{equation}
  for energy density and pressure of a scalar field
 (\ref{e}) and (\ref{p}). Then the next maximum rigid equation of state is valid:
\begin{equation}\label{e=p}
p=\varepsilon,
\end{equation}
and Einstein equation in this case is easily solved and we get:
\begin{equation}\label{a|p=e}
a=a_1\ t^{1/3};\quad \Phi=\phi_0\ln|t|+C_0
\end{equation}
-- $a_1,\phi_0,C_1$ are arbitrary constants where $\phi_0\not=0$.

Let us make the next iteration substituting once again scale factor into source equation of scalar field
\begin{equation}\label{EqF1/3}
\ddot\Phi+\frac{1}{t}\dot\Phi+m^2\Phi=0.
\end{equation}
Solving this equation with an account of massive term:
\begin{eqnarray}\label{Fae2=1}
\Phi=C_1\mathrm{J}_0(m t)+C_2\mathrm{Y}_0(m t),
\end{eqnarray}
where $\mathrm{J}_0(z),\mathrm{Y}_0(z)$ are Bessel functions of actual argument.

Using asymptotics of these functions at small values of argument $z=m t\ll 1$, we obtain from (\ref{Fae2=1}):
\begin{eqnarray}
\Phi\approx C_1 +C_2\frac{2}{\pi}\ln\frac{mt}{2},
\end{eqnarray}
i.e. at small values $m t$ we again get solutions (\ref{a|p=e}). Therefore solutions (\ref{a|p=e}) are also asymptotic solu\-tions of scalar field and Einstein equations obtained at same condition (\ref{approxM2}).
However, cosmological si\-tua\-tion changes drastically. In first case we have an infla\-ti\-onary solution (acceleration $\Omega=1$) with singularity in infinitely distant past while in the second case there exists solution with negative acce\-leration $\Omega=-2$ and singularity in zero instant $t=0$. Situation is completely defined by a choice of constant in first-order equation (\ref{S_Eq_m0}).
This solution of standard scenario corresponding to constant value of scalar field (\ref{F0}) we will further call for brevity sake a {\it standard solution}.

\begin{thm}
Standard cosmological scenario with inflationary beginning corresponds to particular so\-lu\-tion of massless equation of scalar field.
 If using general solution of this equation inflation disappears and the Universe has a finite history and its early stage corresponds to extremely rigid state equation.
\end{thm}

This let us suggest instability of the standard scenario.

\subsection{Stability of Standard Scenario}
Let us consider perturbations of the standard model depending only on time and caused by factor of mass $m$ which we will assume a first order of smallness:
\begin{eqnarray}\label{pert_a}
a(t)=a_0(t)(1+\delta(t));\\
\label{pert_f0}
\Phi(t)=\Phi_0+\phi(t).
\end{eqnarray}
Substituting (\ref{pert_f0}) into equation of scalar field (\ref{Fn0}) and expanding the obtained equation in series by smallness of $\phi,\delta,m^2$, we get the next equation:
\begin{equation}\label{dF}
\ddot{\phi}+3H_0\dot{\phi}+m^2\Phi_0 =0,
\end{equation}
solving which we find (to for simplify writing we put $\Phi_0>0$):
\begin{eqnarray}
\phi=C_1+C_2\mathrm{e}^{-3H_0t}-\Phi_0\frac{m}{\sqrt{3}}t,\nonumber
\end{eqnarray}
where we should put $C_1=0$ since account of this term only brings us to redefining of $\Phi_0$.
Òàêèì îáðàçîì:
\begin{eqnarray}\label{dfol}
\phi=C_2\mathrm{e}^{-3H_0t}-\Phi_0\frac{m}{\sqrt{3}}t.
\end{eqnarray}
Einstein equation's perturbation in approximation linear in $\delta$ brings to equation:
\begin{equation}\label{pert_einst0}
\dot\delta=\frac{4\pi}{3H_0}\delta\varepsilon.
\end{equation}
It is seen from (\ref{e}) that perturbation of energy density is quadratic by disturbances $\phi$ and $m^2$.
Subs\-tituting solution (\ref{dfol}) into equation (\ref{pert_einst0}) and carrying out elementary integration we see that solution has a following form:
\begin{eqnarray}
\delta(t)=A_1 \mathrm{e}^{-6H_0t}+\nonumber\\
\mathrm{e}^{-3H_0t}(A_2+A_3t+A_at^2)+A_5t^3,
\end{eqnarray}
where $A_i$ are certain specific non-zero constants values of which is not important. The important one is the circumstance that in the beginning of the Universe (\ref{begin}), i.e. at $t\to-\infty$ the principle term in the solution is the first one:
\begin{equation}\label{delta-8}
\delta \backsimeq A_1 \mathrm{e}^{-6H_0t}\to \infty,\quad (t\to-\infty) .
\end{equation}
Thus in finite past of the Universe relative per\-tur\-ba\-tion of scale factor caused by the account of massive term in equation of scalar field tends to infinity.
This means instability of the inflationary solution relative to mass factor.
\begin{thm}
Inflationary solution of Einstein\\ equa\-tions (\ref{p=-e}) -- (\ref{inflat}) is instable relative to mass factor:
perturbations of metrics and scalar field grow to infinity in the infinitely distant past.
\end{thm}
The last statement means that infinitely distant past merely did not ever exist.

Above said certainly does not relate to model with cosmological $\Lambda$-term.

\section{Gravitational Perturbations of the Isotropic Universe}
\subsection{Îáùèå ñîîòíîøåíèÿ}
Usually models with primordial inflation are con\-si\-dered as follows:
\begin{equation}\label{week_pert}
g_{ik}=(g^0_{ik}+a^2h_ik)dx^idx^k,
\end{equation}
where $g^0_{ik}$ is a Friedmann metric, $h_{ik}$ are gravitational perturbations.
Herewith usually from {\it aesthetic con\-si\-de\-rations} it is supposed that the Universe beginning is vacuum one with
 state equation $\varepsilon+P=0$,
and gravitational perturbations are small ($h_{ik}\ll 1$), however, the question about the instant when they are small is usually omitted \cite{star}.
For our purposes it is enough to investigate lateral gravitational per\-tur\-bations i.e. gravity waves.

Let us rewrite the metrics with gravitational perturbations in form:
\begin{eqnarray}
\label{Freed}
ds^2_0=a^2(\eta)(d\eta^2-dx^2-dy^2-dz^2);\\
\label{metric_pert}
ds^2=ds^2_0+a^2(\eta)h_{\alpha\beta}dx^\alpha dx^\beta;\\
\label{h(eta)}
h_{\alpha\beta}=e_{\alpha\beta}S(\eta)\mathrm{e}^{i\mathbf{nr}},
\end{eqnarray}
where $S(\eta)$ is an amplitude of gravity waves. Further:
\begin{eqnarray}\label{defh1}
 h^\alpha_\beta=h_{\gamma\beta}g^{\alpha\gamma}_0\equiv-\frac{1}{a^2}h_{\alpha\beta};\\
 \label{defh2}
h\equiv h^\alpha_\alpha\equiv  g^{\alpha\beta}_0h_{\alpha\beta},
\end{eqnarray}
where for gravity waves
\begin{eqnarray}
\label{perp}
h^\alpha_\beta n_\alpha=0;\\
\label{perp1}
h=0.
\end{eqnarray}
As a result of  (\ref{perp1}) in linear in $h$ approximation:
\begin{equation}\label{g}
\sqrt{-g}\approx \sqrt{-g_0}=a^4.
\end{equation}
According to generic approach we consider equation for massless scalar field ($m=0$):
\begin{equation}\label{basecalar}
\Phi''+2\frac{a'}{a}\Phi'-\partial_{\alpha\alpha}\Phi=0
\end{equation}
($'\equiv \partial_\eta$) and its unperturbed solution (\ref{F0}). This, as a result of (\ref{g}) in linear approximation gravity waves do not impact on scalar field. Let us write out formulas for inflationary solution of Einstein equation in this case using time variable $\eta$:
\begin{eqnarray}
a=\mathrm{e}^{\Lambda t}\Rightarrow d\eta=\frac{dt}{a(t)}\Rightarrow\nonumber\\
\label{a(eta)}
a(\eta)=-\frac{1}{\eta},\quad \eta<0;
\end{eqnarray}
In this time variable cosmological singularity $a(\eta)=0$ is matched by value $\eta_{0}=-\infty$, and for infinitely distant future it is $\eta_{+\infty}=-0$ :
\begin{eqnarray}\label{-8}
a(\eta_0)=0\Leftrightarrow \eta_0=-\infty;\\
\label{+8}
a(\eta_\infty)=\infty \Leftrightarrow \eta_\infty=-0.
\end{eqnarray}
%
\subsection{Equation for Aimplitude of Gravity Waves}
Assuming, for certainty::
\begin{equation}\label{nz}
\hskip -12pt \mathbf{n}=n\delta^\alpha_3;\ h_{12}=0; \ h_{11}=-h_{22}=S(\eta)\mathrm{e}^{inz},
\end{equation}
and expanding Einstein tensor in smallness of gravity wave amplitude $S$, we get the following unique non-trivial components in approximation linear in $S$:
\begin{eqnarray}\label{dG}
\delta G_{11}=-\delta G_{22}=\frac{1}{2}\mathrm{e}^{i\mathbf{nr}}\times\nonumber\\
\biggl[S''+2\frac{a'}{a}S'+S\biggl(n^2+2\frac{a'^2}{a^2}-4\frac{a''}{a}\biggr)\biggr]
\end{eqnarray}

Sunstituting expression for potential (\ref{inflat}) in \\ expression (\ref{Tik}) for tensor of scalar field's energy momentum, we find in linear approximation:
\begin{equation}\label{Tik}
T^{(s)}_{\alpha\beta}=\frac{1}{8\pi}m^2a^2h_{\alpha\beta}\Phi^2_0.
\end{equation}
Hence in linear approximation with an account of inflationary solution (\ref{a(eta)}) we get equation on am\-pli\-tude of gravity waves:
\begin{equation}\label{GW_Eq}
S''+\frac{2}{\eta}S'+\biggl(n^2-\frac{6}{\eta^2}-\frac{6H^2_0}{\eta^2}\biggr)S=0,
\end{equation}
solving which we find:
\begin{equation}\label{GWolve}
S(\eta)=\frac{C_1}{\sqrt{\eta}}\mathrm{Y}_\mu(|n|\eta)+\frac{C_2}{\sqrt{\eta}}\mathrm{J}_\mu(|n|\eta),
\end{equation}
where:
\begin{equation}\label{mu}
\mu=\frac{1}{2}\sqrt{25+24H^2_0}.
\end{equation}
In particular, near cosmological singularity it is $\eta\to-\infty$, hence $|\mathbf{n}\eta|\to \infty$
equation (\ref{GW_Eq}) is reduced to equation:
\begin{equation}\label{neta8}
S''+n^2S=0,
\end{equation}
having ordinary WKB-solutions:
\begin{equation}\label{WKBol}
S=C_1e^{in\eta}+C_2e^{-in\eta},
\end{equation}
which, in their turn, can be obtained also from the exact solution (\ref{GWolve}) in this limit.

\subsection{Energy-Momentum of Gravity Waves}
Calculating Landau-Lifshitz pseudotensor \cite{Land_Field} and averaging it in the directions, we obtain for energy-momentum of gravity waves:
\begin{eqnarray}
\label{tik}
\hskip -12pt\mathfrak{T}^{ik}=\frac{1}{a^2}(\mathcal{E}_{gw}+P_{gw})\delta^i_4\delta^k_4-P_{gw}g^{ik},\\
\label{Egw}
\hskip -12pt\mathcal{E}_{g\!w}=-\frac{1}{16\pi}\frac{S^2 n^2}{a^2}-\frac{3}{32\pi}\frac{S'\ \!^2}{a^2}+\frac{5}{8\pi}\frac{SS'a'}{a^3};\\
\label{Pgw}
\hskip -12pt P_{g\!w}=\frac{1}{96\pi}\frac{S^2n^2}{a^2}+\frac{13}{12\pi}\frac{SS'a'}{a^3}-\frac{17}{12\pi}\frac{S^2 a'\ \!^2}{a^4},
\end{eqnarray}
and also:
\begin{eqnarray}\label{E-3Pgw}
\mathfrak{T}=\mathcal{E}_{g\!w}-3P_{g\!w}=-\frac{3}{32\pi}\frac{S'\ \!^2+S^2n^2}{a^2}\nonumber\\
-\frac{21}{8\pi}\frac{SS'a'}{a^3}+\frac{17}{4\pi}\frac{S^2a'\ \!^2}{a^4}.
\end{eqnarray}
Substituting in these formulas inflationary solution (\ref{a(eta)}), we find:
\begin{eqnarray}
\label{Egw_inflat}
\hskip -12pt \mathcal{E}_{g\!w}=-\frac{1}{32\pi}\eta^2(3S'\ \!^2+2S^2n^2)-\frac{5}{8\pi}SS'\eta;\\
\label{Pgw_inflat}
\hskip -12pt P_{g\!w}=\frac{1}{96\pi}S^2n^2\eta^2-\frac{13}{12\pi}SS'\eta-\frac{17}{12\pi}S^2,
\end{eqnarray}
Thus, neat cosmological singularity $\eta\to\-\infty$ we find from (\ref{Egw_inflat}) -- (\ref{Pgw_inflat}):
\begin{eqnarray}
\label{Egw_inflat_0}
\mathcal{E}_{g\!w}\backsimeq -\frac{1}{32\pi}\eta^2(3S'\ \!^2+2S^2n^2);\\
\label{Pgw_inflat_0}
 P_{g\!w}\backsimeq\frac{1}{96\pi}S^2n^2\eta^2,
\end{eqnarray}
Substituting here solutions (\ref{WKBol}), we find:
\begin{equation}\label{ultra_GW}
P_{g\!w}\thickapprox \frac{1}{3}\mathcal{E}_{g\!w}\thickapprox \frac{1}{96\pi}S^2n^2\eta^2.
\end{equation}

\begin{thm}
Thus near cosmological singularity energy density of ``weak'' gravity waves tend to infinity with ultrarelativistic state equation.
 Hence, according to results of chapter 2, the beginning of such Universe is ultrarelativistic rather than inflationary one and
 the Universe itself has the beginning in finite logarith\-mical past (\ref{tg8}). Therefore the model of the Universe with primordial inflation is instable even with respect to weak lateral gravitational perturbations.
\end{thm}
Since perturbation of scalar field caused by gravity waves is absent , above mentioned relates also to models of primordial inflation based on Einstein equations with cosmological term.

Let us evaluate using formulas from Chapter 1 time of cosmological singularity epoch assuming that apart from scalar field in the early stages of  the Universe there are presenet only lateral gravitational perturbations. According to (\ref{a(eta)}) scale factor $a(\eta)$ becomes equal to 1 at $\eta=-1$, but then according to (\ref{gamma}), (\ref{WKBol}) and (\ref{Egw_inflat_0})
\begin{equation}\label{gamma_gw}
\gamma^2=\frac{\pi}{12H^2_0}\overline{n^2S^2(n)},
\end{equation}
where $\overline{n^2S^2(n)}$ is an average value over spectrum of gravitational perturbations. Thus, in such a scenario the beginning of the Universe (cosmological singularity) corresponds to time
\begin{equation}\label{tau_gw}
\tau_{-\infty}=-\frac{1}{2}\ln\frac{12H^2_0}{\overline{n^2S^2(n)}}.
\end{equation}
In such a case the Universe comes to inflationary mode at times
\begin{equation}\label{inflat_trans}
t_{inflat}>t_c=\frac{1}{2H_0}.
\end{equation}
At $S\sim 10^{-10}$, $n\sim 10$, $H_0\sim 10^{-4}$ we obtain from (\ref{tau_gw}) an estimate $\tau_{-\infty}\sim -6\cdot 10^4 t_{pl}$,
where according to (\ref{inflat_trans}) $t_{inflat}\sim 10^4t_{pl}$. Let us remind that this time is calculated from cosmological sin\-gu\-larity i.e.
from standpoint of standard model's timescale it lies in range of negative values.

\section*{Conclusion}
Thus we have shown that model of the Universe with primordial infaltion are physically instable. This instability appears in three forms:

1. Models with early inflation are instable with respect to additions of small additives of physical matter with state equation different from inflationary one $\varepsilon+P=0$. In this case the Universe {\it acquires} its beginning in finite pase and scale factor in the beginning if the Universe history grows according to power law which is then gradually transformed to exponential one;

2. Models with early inflations are instable with respect to degenerated solution with constant scalar field. At correct solution of field equations there appears the Universe with finite beginning;

3.  Models with early inflation are instable with respect to lateral gravitational perturbations which in the early stages provide ultrarelativistic addition to energy-momentum which brings to elimination of the de Sitter stage of evolution.

Here it is necessary to notice that inflation which is early enough (with the beginning of order of $10\div 10^4$ Planck times) takes place (see e.g. \cite{Yu_LTE,Yu_Ass}). Let us notice that according to contemporary observations for solution of horizon and planeness problems it is enough to have inflation duration $10^{-42}\div 10^{-9}$ s (see e.g. \cite{Gorb_Rubak}), i.e. post-Planck inflation is enough.
Apparently, stage preceding early inflation is the ultrarelativistic stage generated by massless fields.

\section{Acknowledgements}

In conclusion Author would like to express his gratitude to participants of seminar for relativistic kinetics and cosmology (MW) of Kazan Federal Uni\-ver\-sity for helpful discussion of the work.

\end{document}